\documentclass{PoS}

\usepackage{amssymb}
\usepackage{fontenc}
\usepackage{times}
\usepackage{mathptmx}
\usepackage{graphicx}
\usepackage{bbm}
\usepackage{amsmath}

\title{A test of Taylor- and modified Taylor-expansion}

\ShortTitle{A test of Taylor- and modified Taylor-expansion}

\author{\speaker{Max Wilfling}\\
Institute of Physics, Karl-Franzens Universit\"at Graz, Austria\\
E-mail: \email{maximilian.wilfling@edu.uni-graz.at}}

\author{Christof Gattringer\\
Institute of Physics, Karl-Franzens Universit\"at Graz, Austria\\  
 E-mail: \email{christof.gattringer@uni-graz.at}}

\abstract{We compare Taylor expansion and a modified variant of Taylor expansion, which incorporates 
features of the fugacity series, for expansions in the chemical potential around a zero-density lattice field theory. As a first test we apply both series 
to the cases of free fermions and free bosons.
Convergence and other properties are analyzed.}

\FullConference{31st International Symposium on Lattice Field Theory LATTICE 2013\\
July 29 - August 3, 2013\\
Mainz, Germany}

\begin{document}

\section{Introduction}
At non-zero chemical potential $\mu$ the fermion determinant becomes complex and 
cannot be used as a probability weight in a Monte Carlo simulation of finite density lattice QCD.
This so-called ''complex action problem'' (or ''sign problem'') has been  a major obstacle on the 
way to an ab-initio treatment of QCD at finite $\mu$ on the lattice. For small $\mu$ a possible way 
out are various series expansions where the coefficients of the expansion can be computed 
in a simulation at $\mu = 0$. The simplest expansion is the Taylor series in $\mu$ (see, e.g., \cite{taylor}), 
with the advantage that the coefficients can be evaluated with standard techniques. 
An interesting alternative is fugacity expansion, which on a finite lattice is a finite Laurent series in the fugacity 
parameter $e^{\beta \mu}$. For the fugacity series the expansion coefficients become small quickly,  
but on the other hand are very costly to compute as Fourier moments 
of the fermion determinant with respect to imaginary chemical potential \cite{fugacity}. 

In this contribution we analyze regular Taylor expansion (RTE) and a 
modified Taylor expansion (MTE) and compare the results to the exact 
results from Fourier transformation using both free fermions and free bosons. The MTE is an expansion in 
\begin{equation}
\rho=e^\mu-1\quad , \quad \overline{\rho}=e^{-\mu}-1 \; ,
\end{equation}
which for small $\mu$ reduces to the conventional Taylor expansion. On the other hand it captures 
features of the fugacity series, but the coefficients 
come with the same price tag as the coefficients of the Taylor series. A comparison \cite{Z3} of the regular 
and the modified Taylor expansions in a toy model, the $\mathbbm{Z}_3$ center model, showed 
that the MTE outperforms the RTE for a wide range of parameters. 

\section{The modified Taylor expansion MTE}
In this work we use free fermions, as well as free bosons at finite density. The corresponding lattice actions 
are given by  
\begin{equation}
S_F(\mu) \;  = \; \sum_{x} \left[ (m + 4)  \overline{\psi}_x  \psi_x - \sum_{\nu = 1}^4 \left[ e^{\mu \delta_{\nu,4}}
 \overline{\psi}_x \frac{\mathbbm{1} - \gamma_\nu}{2} \psi_{x+\hat{\nu}} +  e^{-\mu \delta_{\nu,4}}\overline{\psi}_x \frac{\mathbbm{1} + \gamma_\nu}{2} \psi_{x-\hat{\nu}} \right] \right] \; ,
\label{sfermion}
\end{equation}
for the fermions (we use the Wilson action), and by
\begin{equation}
S_B(\mu) \;  = \; \sum_{x} \left[ (m^2 + 8)  \phi^*_x  \phi_x - \sum_{\nu = 1}^4 \left[ e^{- \mu \delta_{\nu,4}}
\phi^*_x \phi_{x+\hat{\nu}} +  e^{+\mu \delta_{\nu,4}} \phi^*_x \phi_{x-\hat{\nu}}   \right] \right] \; ,
\label{sboson}
\end{equation}
for the bosons.
The fields live on the sites of $N_s^3 \times N_t$ lattices and we use periodic boundary conditions for all directions, except for
the temporal direction in the fermionic case where the boundary conditions are anti-periodic. Throughout this paper the lattice spacing is
set to $a=1$ and all results are in lattice units. For later use we define the 3-volume in lattice units as $V = N_s^3$ and the inverse temperature
in lattice units as $\beta = N_t$ (the Boltzmann constant is set to $k_B = 1$).

For both the fermion and boson cases we decompose the action as
\begin{equation}
S(\mu) \; = \; S(0) \; - \; \rho R \; - \; \overline{\rho} \overline{R} \;,
\end{equation}
where for the fermions
\begin{equation}
R \; = \; \sum_{x} \overline{\psi}_x \frac{\mathbbm{1} - \gamma_4}{2} \psi_{x+\hat{4}} \quad , \qquad \overline{R} \; = \; \sum_{x} \overline{\psi}_x \frac{\mathbbm{1} + \gamma_4}{2} \psi_{x-\hat{4}} \; ,
\label{rfermion}
\end{equation}
and
\begin{equation}
R \; = \; \sum_{x} \phi^*_x  \, \phi_{x-\hat{4}} \quad , \qquad \overline{R} \; = \; \sum_{x} \phi^*_x  \, \phi_{x+\hat{4}} \; ,
\label{rboson}
\end{equation}
for the bosons. The partition sum can then be written as
\begin{equation}
Z(\mu) \; = \; Z(0) \, \left\langle e^{\, \rho R} \, e^{\, \overline{\rho} \overline{R}} \right\rangle_0 \; ,
\end{equation}
where $\langle .. \rangle_0$ denotes the expectation value at $\mu = 0$. Series expansion of the two exponentials leads to the MTE
\begin{equation}
Z(\mu) \; = \; Z(0) \, \sum_{n, \overline{n} = 0}^\infty \frac{ \rho^n \; \overline{\rho}^{\; \overline{n}}}{ n! \, \overline{n} ! } \, \left\langle R^{\, n} \, \overline{R}^{\; \overline{n}} \right \rangle_0 \; .
\label{mte}
\end{equation}
We remark that it is straightforward to include gauge fields in the formalism by simply adding the gauge links $U_\nu(x)$ in the nearest neighbor terms 
in (\ref{sfermion}), (\ref{sboson}), (\ref{rfermion}), (\ref{rboson}) -- all other expressions remain the same.  

In addition we also consider the conventional Taylor expansion of the partition sum,
\begin{equation}
Z(\mu) \; = \; \sum_{n=0}^\infty c_{2n} \; \mu^{2n}  \quad , \qquad c_{2n} \; = \; \frac{1}{2n !} 
\left(\frac{\partial}{\partial \mu}\right)^{\!\!2n} \, Z(\mu) \Big|_{\mu = 0} \; .
\label{rte}
\end{equation}

For both expansions we study as our observables the free energy density $f$, the particle number density $n$ and the particle number susceptiblity 
$\chi_n$ which were calculated according to
\begin{equation} 
f  \; = \; -\frac{1}{V\beta}\ \textrm{ln}\ Z(\mu) \; , \; \; 
n\; = \; \frac{1}{V} \frac{\partial}{\partial(\mu\beta)}\ \textrm{ln}\ Z(\mu) \; , \; \;
\chi_n \; = \; \frac{1}{V} \frac{\partial^2}{\partial(\mu\beta)^2}\ \textrm{ln}\ Z(\mu)  \; .
\label{obs}
\end{equation}
Of course the expansions (\ref{mte}) and (\ref{rte}) and thus also the resulting series (\ref{obs}) for the observables have to be truncated. This is done 
by truncating the series for $\textrm{ln}\ Z(\mu)$ (obtained by inserting the series for $Z(\mu)$ into the logarithm) at the desired order in $\mu^2$
or in $\rho$ and $\overline{\rho}$.   
 
\section{Solution of the free case from Fourier transformation}

For the free case which we consider here, the partition sum can be computed in closed form with the help of Fourier transformation. The momenta
are given by 
\begin{equation}
p_i \, = \; \frac{2\pi}{N_s} n_i \; , \; n_i = 0,1,2 \, ... \, N_s - 1 \;\;\; \mbox{for} \;\;\; i = 1,2,3 \; , \quad  
p_4 \, = \; \frac{2\pi}{N_t} (n_4 + \theta) \; , \; n_4 = 0,1,2 \, ... \, N_t - 1 ,
\end{equation}
where $\theta = 1/2$ for the fermions and $\theta = 0$ for bosons. Below we display the exact results for $\ln Z(\mu)$ 
for both the fermionic and the bosonic case, as functions of $\rho$ and $\overline{\rho}$. From that one can immediately 
obtain the MTE for $\ln Z(\mu)$ by expansion in $\rho$ and $\overline{\rho}$ and the RTE by expansion in $\mu^2$. Subsequently the
observables were computed with the help of (\ref{obs}).

\vskip5mm
\noindent
{\bf Fermions:}
\\
Using standard techniques (see, e.g., \cite{gala}), the logarithm of the canonical partition function $\ln Z(\mu)$ for fermions is obtained in closed form as:
\begin{equation}
 \textrm{ln}\ Z(\mu) \; = \; 2\sum\limits_p \bigg[ \, \textrm{ln}\ R_p \, + \, \textrm{ln}\ (\, 1 \, - \, a_p \, \rho \; - \; a_p^* \,
 \overline{\rho} \, ) \,  \bigg], 
 \label{lnzf}
\end{equation}
with

\begin{equation}
a_p \; = \; \frac{c_p\ e^{ip_4}}{R_p},\ R_p \; = \; c_p^2-2\ c_p\ \textrm{cos}p_4+\sum\limits_{i=1}^3 \textrm{sin}^2p_i+1,\ c_p \; = \; m+4-\sum\limits_{i=1}^3 \textrm{cos}p_i \; .
\end{equation}

\vskip5mm
\noindent
{\bf Bosons:}
\\
For the bosonic case one obtains for $\ln Z(\mu)$ in closed form the expression:
\begin{equation}
 \textrm{ln}\ Z(\mu) \; = \; V\beta\ \textrm{ln}\ 2\pi \; -\; \sum\limits_p \bigg[ \textrm{ln}\ R_p \; + \;
 \textrm{ln}\ ( 1 \; - \; a_p \, \rho \; - \; a_p^* \, \overline{\rho}) \bigg], 
 \label{lnzb}
 \end{equation}
where
\begin{equation}
a_p \; = \; \frac{e^{i p_4}}{R_p} \; , \; 
R_p \; = \; m^2+8 \, - \, 2\sum\limits_{i=1} ^3 \textrm{cos}p_i \, - \, 2\ \textrm{cos}p_4 \; .
\end{equation}

\vskip3mm
\noindent
In both cases one can now perform an expansion of $\ln Z(\mu)$ in  $\rho$ and $\overline{\rho}$ to obtain the MTE, or an 
expansion in $\mu^2$ for the RTE. The expansions are then truncated at some order and subsequently we compute the 
derivatives necessary for the observables (\ref{obs}). For comparison we use the exact results for the observables, i.e., we 
compute the observables directly from (\ref{lnzf}) and (\ref{lnzb}) without any expansion.

\section{Results} 

In this section we now present the observables (\ref{obs}) from the MTE and the RTE at different orders for the truncation and compare the results 
to the exact expressions. The results we show here were obtained on $32^3 \times 4$ lattices with a fermion mass of $m=0.1$ and for the bosons 
with a mass of $m = 1.0$. Tests with other masses and volumes gave rise to comparable results.

\begin{figure}[p]\begin{center}
\includegraphics[width=140mm]{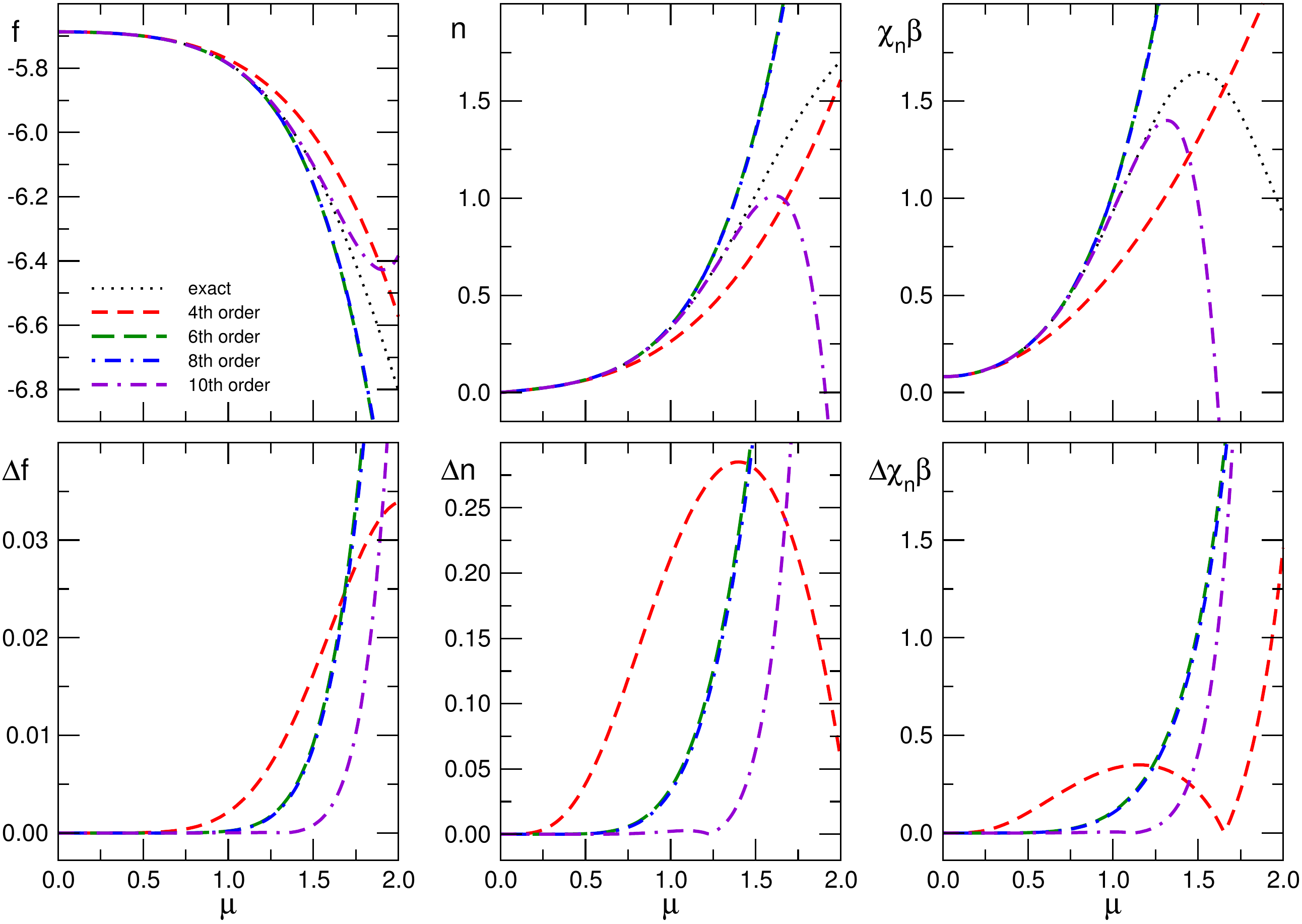}
\caption{Assessment of the RTE in the fermionic case: We show our observables (top row of plots) and relative errors (bottom) 
as a function of $\mu$.}
\label{rte_fermions}
\end{center}
\begin{center}
\includegraphics[width=140mm]{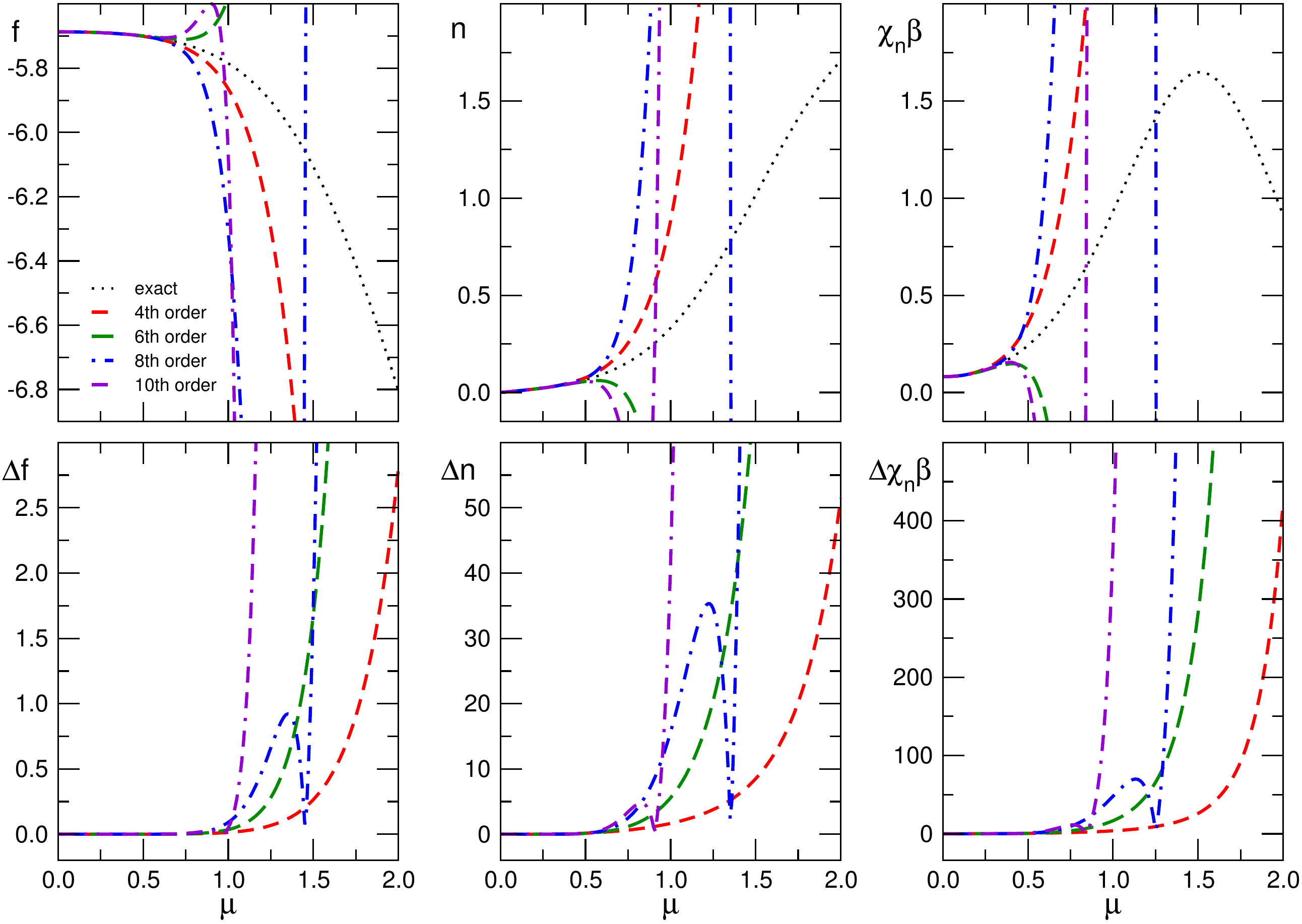}
\caption{Assessment of the MTE in the fermionic case: We show our observables (top row of plots) and relative errors (bottom) 
as a function of $\mu$.}
\label{mte_fermions}
\end{center}
\end{figure}

We begin with the discussion of the results for the fermionic case. In the top row of plots in 
Fig.~\ref{rte_fermions} we show the observables $f$, $n$ and 
$\chi_n \beta$ as a function of $\mu$. We display the results from the RTE taking into account terms up to orders ${\cal O}(\mu^4)$,
 ${\cal O}(\mu^6)$,  ${\cal O}(\mu^8)$ and  ${\cal O}(\mu^{10})$ and compare them to the exact result. In the bottom row of plots we show 
 the corresponding relative errors. We observe that  when including terms up to order ${\cal O}(\mu^{10})$ in the RTE, one finds a good  
 representation of the exact result up to $\mu = 1$ for all our observables. Truncating at lower orders reduces this range to smaller values of $\mu$.

\begin{figure}[p]\begin{center}
\includegraphics[width=140mm]{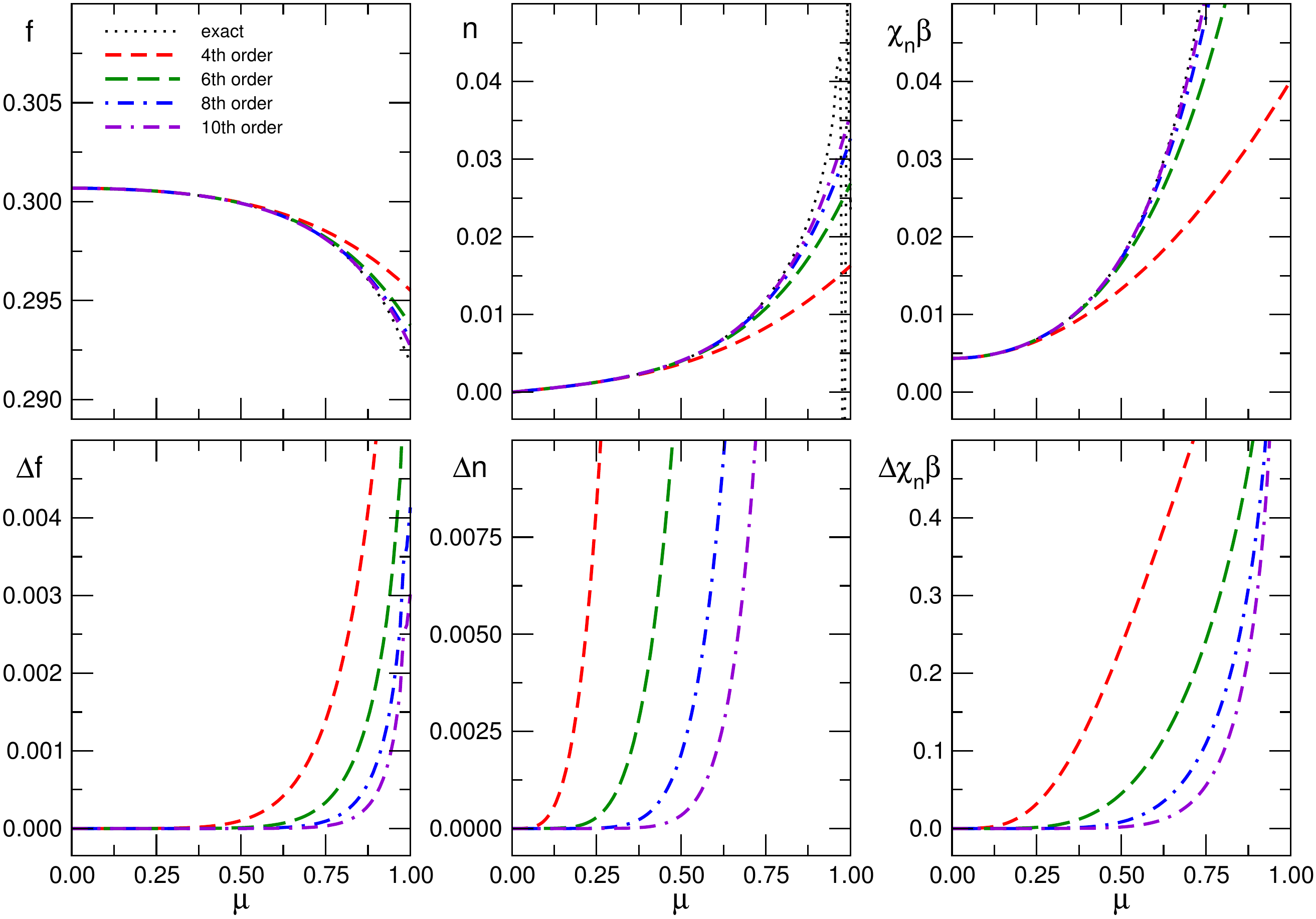}
\caption{Assessment of the RTE in the bosonic case: We show our observables (top row of plots) and relative errors (bottom) 
as a function of $\mu$.}
\label{rte_bosons}
\end{center}
\begin{center}
\includegraphics[width= 140mm]{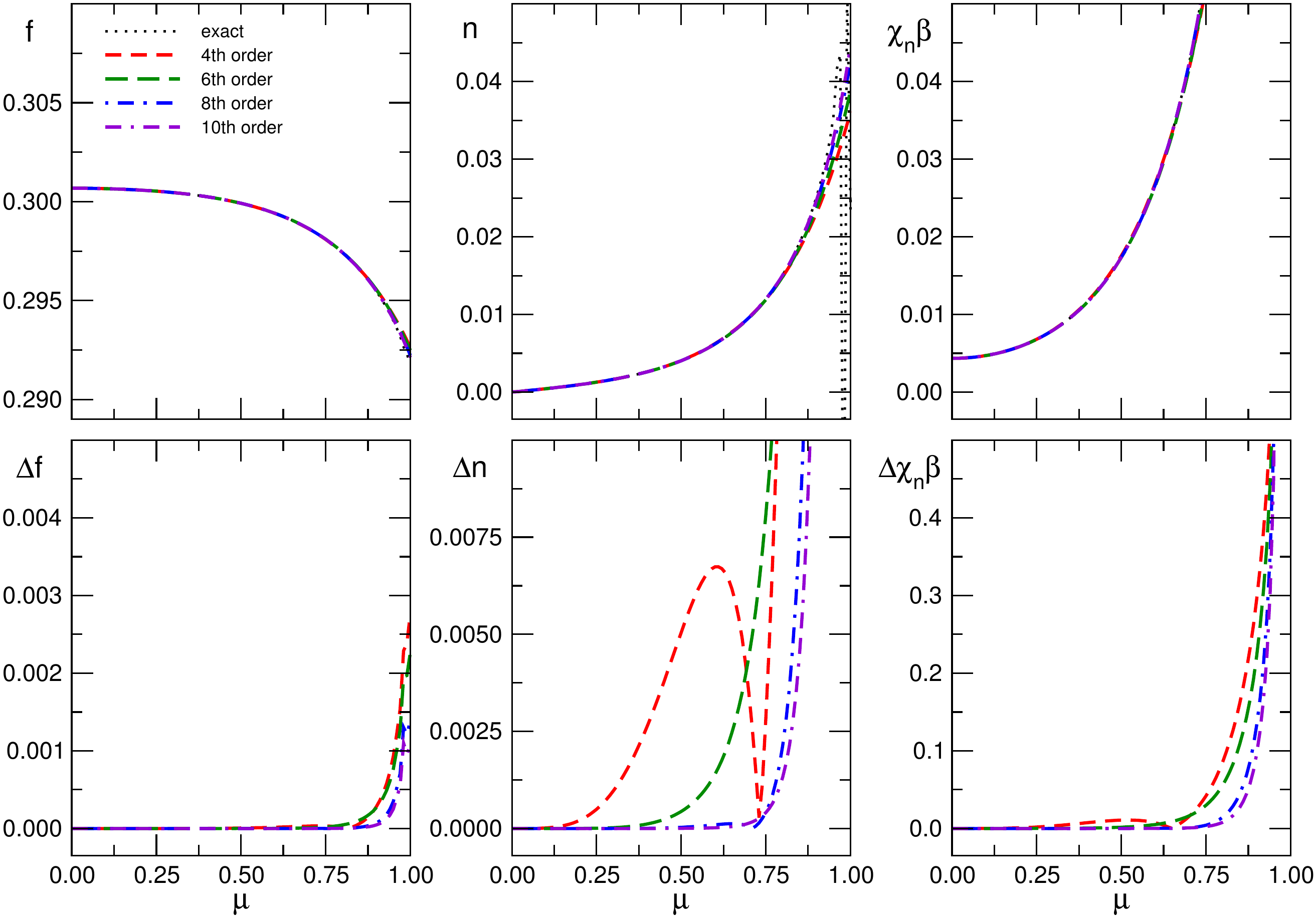}
\caption{Assessment of the MTE in the bosonic case: We show our observables (top row of plots) and relative errors (bottom) 
as a function of $\mu$.}
\label{mte_bosons}
\end{center}
\end{figure}

In Fig.~\ref{mte_fermions} we present the same observables as in Fig.~\ref{rte_fermions}, but now for the modified Taylor expansion
MTE. In this 
case we find only a smaller range in $\mu$ where the ${\cal O}(\mu^{10})$ series is reliable -- roughly up to $\mu \sim 0.6$. 

In Fig.~\ref{rte_bosons} and Fig.~\ref{mte_bosons} we repeat the analysis of Fig.~\ref{rte_fermions} and Fig.~\ref{mte_fermions}, now for the 
case of bosons. Here the role of MTE and RTE is reversed with a slightly better representation of the exact results by the MTE, which gives a
reliable representation up to $\mu \sim 0.75$.
 
\section{Discussion and remarks}

In this project we explore a new variant of the Taylor expansion, MTE, and compare it to the regular Taylor expansion, RTE, 
for free bosons and free fermions. The MTE is an expansion in $e^{\pm \mu} - 1$ and thus combines properties of the
fugacity expansion (a Laurent series in $e^{\beta \mu}$) with aspects of the Taylor series. 

We find that in the fermionic case the RTE is the most viable method for approximating the exact results, with the MTE being less accurate
for the parameters we studied.  For bosons the situation is reversed and the 
MTE proves to be the more precise method. These results are somewhat unexpected as 
in a recent comparison of the RTE and the MTE in an effective theory for the center degrees of freedom of QCD \cite{Z3} it was found that 
the MTE very clearly outperforms the RTE. Why this is not the case in the free theories studied here will have to be the subject of future studies.


\begin{thebibliography}{99}

\bibitem{taylor}
  C.R.~Allton, S.~Ejiri, S.J.~Hands, O.~Kaczmarek, F.~Karsch, E.~Laermann, C.~Schmidt,
  %``The Equation of state for two flavor QCD at nonzero chemical potential,''
  Phys.\ Rev.\ D 68, 014507 (2003).
  %[hep-lat/0305007].
  %%CITATION = HEP-LAT/0305007;%% 
  C.R.~Allton, M.~Doring, S.~Ejiri, S.J.~Hands, O.~Kaczmarek, F.~Karsch, E.~Laermann, K.~Redlich,
  %``Thermodynamics of two flavor QCD to sixth order in quark chemical potential,''
  Phys.\ Rev.\ D 71, 054508 (2005).
  %[hep-lat/0501030].
  %%CITATION = HEP-LAT/0501030;%% 
 
\bibitem{fugacity}
J.~Danzer, C.~Gattringer,
  %``Properties of canonical determinants and a test of fugacity expansion for finite density lattice QCD with Wilson fermions,''
  Phys.\ Rev.\ D 86 (2012) 014502.
  %%CITATION = ARXIV:1204.1020;%%
E.~Bilgici, J.~Danzer, C.~Gattringer, C.B.~Lang, L.~Liptak,
  %``Canonical fermion determinants in lattice QCD: Numerical evaluation and properties,''
  Phys.\ Lett.\ B 697 (2011) 85.
  %%CITATION = ARXIV:0906.1088;%%
  
\bibitem{Z3}
E.~Gr\"unwald, Y.~D.~Mercado and C.~Gattringer,
  %``Taylor- and fugacity expansion for the effective center model of QCD at finite density,''
PoS (LATTICE 2013) 448 [arXiv:1310.6520].
  %%CITATION = ARXIV:1310.6520;%%
Y.~Delgado, H.G.~Evertz and C.~Gattringer, 
Phys.\ Rev.\ Lett.\ 106 (2011) 22200;
%%CITATION = ARXIV:1102.3096;%%
Comput.\ Phys.\ Commun.\  183 (2012) 1920.
  %%CITATION = ARXIV:1202.4293;%%

\bibitem{gala}  
 C.~Gattringer and C.~B.~Lang, {\it Quantum chromodynamics on the lattice}, Lecture Notes in Physics 788, Springer (2010).
 %%CITATION = LNPHA,788,1;%%

\end{thebibliography}
\end{document}